\documentstyle[eqsecnum,preprint,aps]{revtex}

\def\footnoterule{\kern-3pt \hrule width \hsize \kern6.2pt}

\def\pmb#1{\setbox0=\hbox{$#1$}%
\kern-.025em\copy0\kern-\wd0
\kern.05em\copy0\kern-\wd0
\kern-.025em\raise.0433em\box0 }

\tighten

\begin{document}

\title{Spherical Shells of Classical Gauge Field\\
and Their Topological Charge as a Perturbative Expansion}

\author{Edward Farhi\footnote[1]{farhi@mitlns.mit.edu.~~
Research supported in part by the U.S.~Department of Energy (D.O.E.)
under contract \#DE-AC02-76ER03069}}
\smallskip
\address{Center for Theoretical Physics\\
Laboratory for Nuclear Science and\\
Department of Physics\\
Massachusetts Institute of Technology\\
Cambridge, MA\ \ 02139}

\author{Valentin V.~Khoze\footnote[2]{
vkhoze@slacvm.slac.stanford.edu. ~~Research supported in part by the D.O.E.
under contract \#DE-AC03-76SF0015.}}
\smallskip
\address{Stanford Linear Accelerator Center\\
Stanford University\\
Stanford, CA\ \ 94309}

\author{Krishna Rajagopal\footnote[3]{rajagopal@huhepl.harvard.edu.~~
Junior Fellow, Harvard Society of Fellows.~~Research supported in part
by the National Science Foundation under grant PHY-92-18167.}}
\smallskip
\address{Lyman Laboratory of Physics\\
Harvard University\\
Cambridge, MA\ \ 02138}

\author{Robert Singleton, Jr.\footnote[4]{bobs@pth3.bu.edu.~~Research
supported in
part by the D.O.E. under contract \#DE-FG02-91ER40676 and
by the Texas National Research Laboratory Commission
under grant RGFY93-278.}}
\smallskip
\address{Department of Physics\\
Boston University\\
Boston, MA\ \ 02215}

\maketitle

\setcounter{page}{0}
\thispagestyle{empty}

\vfill

\noindent CTP\#2287

\noindent SLAC-PUB-6419(T) \hfill March 1994

\noindent HUTP-94-A003 \hfill Submitted to {\it Physical Review} {\bf D}

\noindent BUHEP-94-6 \hfill Typeset in REV\TeX
\eject

\vfill

\begin{abstract}

We consider the classical equations of motion of $SU(2)$ gauge theory,
without a Higgs field, in Minkowski space.
We work in the spherical ansatz and develop a perturbative
expansion in the coupling constant
$g$ for solutions which in the far past look like
freely propagating spherical shells.  The topological charge $Q$ of these
solutions is typically non-integer.  We then show that $Q$
can
be expressed as a power series expansion in $g$ which
can be nonzero at finite order.
We give an explicit
analytic calculation of the order
$g^5$ contribution to $Q$
for specific initial pulses.
We discuss the relation between our findings and anomalous
fermion number violation, and speculate on the
physical implications of our results.
\end{abstract}

\vfill

\eject

\baselineskip 24pt plus 2pt minus 2pt

\section{Introduction}
\label{sec:level1}

Fermion number is not conserved in the
standard model \cite{thooft}, yet reliable techniques for calculating
fermion number violating rates in high energy collisions have
not been fully developed.  At very low energy where the processes
are best described as tunnelling events, the
calculational methods \cite{thooft} make use of solutions
to the Euclidean field equations --- instantons \cite{BPST}.
At energies comparable to but below the sphaleron barrier \cite{manton},
Euclidean methods \cite{ringwaldplus}
and other methods in which  part of the calculation
is done in Euclidean space \cite{rubakov} have also been applied.
However, at energies well above the tunnelling
barrier, it may
be more appropriate to work directly in Minkowski
space \cite{christ,FKS}.  To this end, we are investigating
solutions to the Minkowski space classical equations of
motion.  In this paper, we study
$SU(2)$ gauge theory without the Higgs field.

Working in the (spatial) spherical ansatz, we explore solutions
which  have the property that in the far past they
describe freely propagating incoming shells of
energy.  We do this by developing a
perturbative expansion in the coupling $g$ which
can be used to systematically solve the equations
of motion once arbitrary initial profiles have been specified.
We discuss the topological charge of these solutions,
which we also develop in a power series expansion in $g$.
We show that the topological charge is nonzero at order
$g^5$.  We then investigate the associated anomalous
fermion production
in the presence of these solutions.

Recently \cite{FKS}, it was shown that in the spherical
ansatz the classical equations of motion of  $SU(2)$
gauge theory can be reduced to two coupled nonlinear partial
differential equations of motion for two gauge invariant functions
of
$r=|{\bf
x}|$ and $t$.
In Ref. \cite{FKS}, new solutions to these equations were also presented
which have
much in common with
the previously discovered solutions \cite{LS}
of L\"uscher and Schechter (LS).
As we will show here
these explicit solutions are examples of a wide class of finite energy
solutions all of which have certain general features in common.
At early times they
depict a thin spherical shell of energy imploding towards the origin at near
the speed of light.  At intermediate times the region around the origin is
energetically excited and at late times the shell is expanding outward,
asymptotically approaching the speed of light.

We are particularly interested in solutions which look free in the far past.
This is because our ultimate goal is to connect the classical solutions to the
quantum mechanical description of the scattering of quanta which in the far
past are well separated and non-interacting.  For a classical solution to be
said to be free in the far past we require that there exists a gauge in which
the commutator term in $F_{\mu\nu}$ (see 1.3) can be neglected in the far
past.  This means that the theory linearizes.  We also want $A_\nu$ in this
gauge to be independent of $g$ in the far past.  This means that the initial
value data can be specified without reference to $g$.
The LS solutions and the solutions of Ref. \cite{FKS} do not satisfy
this requirement. However, in this paper
we will show, in the
spherical ansatz, how to specify initial data which do meet this criterion.

Returning to the description of the shells,
if we focus on the region near the origin we see that before the front reaches
it, it is devoid of energy, {\em i.e.\/} it is a vacuum configuration.  The
energy shell excites this region and then leaves so that at late times the
region is once again a vacuum configuration.  However, the late time vacuum
structure of this region need not coincide with its early time structure.
As we now explain, the region can be
associated with a local topological charge which is typically nonvanishing,
non-integer, and $g$-dependent.

We define the local topological charge inside a sphere of radius $R$ as
\begin{equation}
Q(R)=\frac{g^2}{32\pi^2}\int_{-\infty}^{\infty}dt\int_R
d^3x\,\epsilon^{\mu\nu\alpha\beta}{\rm Tr}(F_{\mu\nu}F_{\alpha\beta})\ ,
\label{1.1}
\end{equation}
where the spatial integration is over the interior of the sphere.  For the
LS
solutions, $Q(R)$ approaches a fixed value as $R$ goes to infinity whereas for
the solutions of Ref.
\cite{FKS}, $Q(R)$ grows without bound.  Let us restrict
our attention to all those
solutions for which $Q(R)$ has a limit as $R$ goes to infinity.  In this case,
as $R$ goes to infinity,
$Q(R)$ approaches the topological charge $Q$ defined by
\begin{equation}
Q=\frac{g^2}{32\pi^2}\int
d^4x\,\epsilon^{\mu\nu\alpha\beta}{\rm Tr}(F_{\mu\nu}F_{\alpha\beta})\ ,
\label{1.2}
\end{equation}
where the integration is over all of space-time.

The topological charge
$Q$ can take any value, not just integer values \cite{FKS}.  At first sight
this may appear
to contradict the observation that (\ref{1.2}) is a topological
invariant of the field configuration.  Let us review the usual argument
\cite{J} that
leads to integer topological charge.
Imagine that the region of space-time where the energy density is nonzero is
bounded.  Surround this region with a three dimensional surface which is
topologically $S^3$.  On this surface $F_{\mu\nu}=0$ so the gauge field is
pure gauge, {\em i.e.\/} $A_{\mu}=(i/g)\,U\partial_\mu U^{\dagger}$
where $U$ takes
values in $SU(2)$.  This defines a map from $S^3$ into $SU(2)$ which is
characterized by an integer since $\pi_3(SU(2))= Z $.  The integrand in
(\ref{1.2}) is a total divergence so $Q$ can be written as an integral over
the surrounding three dimensional surface.
This integral computes the winding number of the map from $S^3$ into $SU(2)$
and therefore $Q$ is an integer.

This argument assumes that the space-time region
where the energy density is nonzero is
bounded.  For solutions to the equations of motion, energy is conserved
and the energy computed on {\it any}
equal time surface is non-vanishing.  It is therefore
impossible to surround the space-time
region of nonzero energy density with a three
sphere.  We have no reason to expect and we do not find integer values of $Q$.

If the topological argument applied and $Q$ were an integer then certainly $Q$
would be
independent of the gauge coupling $g$.  We now examine another commonly used
argument which, when applicable, says that
$Q$ is independent of $g$.  The field strength $F_{\mu\nu}$ is defined as
\begin{equation}
F_{\mu\nu}=\partial_\mu A_\nu-\partial_\nu A_\mu-ig[A_\mu,A_\nu]\ ,
\label{1.3}
\end{equation}
and the equation of motion is
\begin{equation}
D^\mu F_{\mu\nu}=\partial^\mu F_{\mu\nu}-ig[A^\mu,F_{\mu\nu}]=0\ .
\label{1.4}
\end{equation}
We can redefine the fields so that the equation of motion is independent of
$g$.  Let
\begin{equation}
A_\mu=\frac{1}{g} {A'}_\mu\ ,
\label{1.5}\end{equation}
\begin{equation}
\phantom{x}\nonumber\\
F_{\mu\nu}=\frac{1}{g} {F'}_{\mu\nu}=\frac{1}{g}\left(\partial_\mu
{A'}_\nu-\partial_\nu{A'}_\mu-i[{A'}_\mu,{A'}_\nu]\right)\ .
\label{1.6}
\end{equation}
Now the equation of motion reads
\begin{equation}
\partial^\mu {F'}_{\mu\nu}-i[{A'}^\mu,{F'}_{\mu\nu}]=0\ ,
\label{1.7}
\end{equation}
which contains no $g$.  The topological charge (\ref{1.2}) is seen to be
independent of $g$:
\begin{equation}
Q[{A'}_\mu]=\frac{1}{32\pi^2} \int
d^4x\, \epsilon^{\mu\nu\alpha\beta}{\rm
Tr}\left({F'}_{\mu\nu}{F'}_{\alpha\beta}\right)\ .
\label{1.8}
\end{equation}
In this sense solutions to the equation of motion yield a topological charge
independent of the coupling.

However, we are interested in solving (\ref{1.4})
in Minkowski space with fixed initial value data
which does not depend on the coupling $g$.  That is, we wish to solve
(\ref{1.4}) specifying the field $A_\mu$ and its time derivative $\dot A_\mu$
at some fixed time $T$.  We specify
\begin{eqnarray}
A_\mu({\bf x},T) &=& A^{\rm init}_{\mu}({\bf x})\ ,
\label{1.9}\\
\dot A_\mu({\bf x},T) &=& V^{\rm init}_{\mu}({\bf x})\ ,
\label{1.10}
\end{eqnarray}
where
$A^{\rm init}_\mu({\bf x})$ and $V^{\rm init}_\mu({\bf x})$ are arbitrary
functions consistent with the constraints.  In terms of rescaled variables
\begin{eqnarray}
{A'}_\mu({\bf x},T) &=& gA^{\rm init}_\mu({\bf x})\ ,
\label{1.11}\\
\dot{A'}_\mu({\bf x},T) &=& gV^{\rm init}_\mu({\bf x})\ .
\label{1.12}
\end{eqnarray}
If we fix $A^{\rm init}_\mu({\bf x})$ and $V^{\rm init}_\mu({\bf x})$ then
${A'}_\mu$ actually depends on $g$.  Accordingly $Q$ depends on $g$.

As an
example, if $g$ is equal to zero, we are solving (\ref{1.7}) with the
condition that ${A'}_\mu$ and $\dot{A'}_\mu$ are zero at $T$.  The solution is
${A'}_\mu({\bf x},t)=0$ which yields zero topological charge in (\ref{1.8}).
Another way to see this is to work with $g=0$ directly.  The equation of
motion (\ref{1.3}) is
\begin{equation}
\partial^\mu(\partial_\mu A_\nu-\partial_\nu A_\mu)=0\ ,
\label{1.13}
\end{equation}
which we can solve with fixed initial value data, yielding non-trivial
solutions.  In this case $Q$ given by (\ref{1.2}) is seen to be zero since $g$
is zero.
Thus if we solve the classical equations of motion with fixed,
$g$-independent, initial value
data we expect the topological charge to vanish as $g$ goes to zero.
This suggests that we can develop a power series expansion
in $g$ for $Q$.

To conclude this introduction, we describe the
organization of the rest of the paper.
In section II we review the spherical
ansatz and write the equations of motion
in terms of  two gauge
invariant variables.
In section III we show that in
the spherical ansatz, the theory linearizes in the far
past and the far future.
We express the requirement that $A^\mu$ be $g$-independent
in the far past in terms of the gauge invariant variables.
In section IV we show how to solve the field
equations as a perturbative expansion in g for arbitrary
initial pulses, and perform the calculation explicitly
to zeroth
and first order in g.
In section V we return to our discussion of the
topological charge of the solutions.  We show how
to expand $Q$ as a power series in $g$.
We present
examples of initial data for which
the topological charges of the corresponding solutions
are nonzero at order $g^5$.
In appendix A, without restricting to the spherical
ansatz, we show that in source-free electromagnetism
the topological charge of localized pulses vanishes.
In section VI we discuss anomalous fermion number
violation in the presence of classical gauge field sources.
We show that the number of fermions produced, in
a quantum average sense, is $Q$.
In section VII, we remark upon
the implications of our results
for anomalous fermion number violation
in high energy scattering
in the Standard Model, and discuss open questions.

\vskip .3in

\section{The Spherical Ansatz}
\label{sec2}

Here we review the spherical ansatz \cite{W}
for $SU(2)$ gauge theory without a Higgs field and show how
the equations can be reduced to two equations for two gauge invariant
functions.
The spherical ansatz is given by expressing the gauge field $A_\mu$ in terms
of four functions $a_0\, ,\,a_1\, ,\,\alpha\ {\rm and}\ \gamma\ {\rm of}\ r\
{\rm and}\ t$:
\begin{eqnarray}
A_0({\bf x},t) &=& \frac{1}{2}a_0(r,t){\pmb\sigma}\cdot{\bf \hat x}
\nonumber\\
A_i({\bf x},t) &=& \frac{1}{2}\big[a_1(r,t){\pmb\sigma}\cdot{\bf\hat x}\hat
x_i+\frac{\alpha(r,t)}{r}(\sigma_i-{\pmb\sigma}\cdot{\bf\hat x}\hat x_i)
+\frac{\gamma(r,t)}{r}\epsilon_{ijk}\hat x_j\sigma_k\big]\ ,
\label{2.1}
\end{eqnarray}
where ${\bf \hat x}$ is the unit three-vector in the radial direction.
Note that
we do not introduce $1/g$ factors in (\ref{2.1}) as
was done in Refs. \cite{W} and \cite{FKS}.
The action
\begin{equation}
S=-\frac{1}{2}\int d^4x\,Tr(F_{\mu\nu}F^{\mu\nu})\ ,
\label{2.2}
\end{equation}
in the spherical ansatz takes the form of an Abelian Higgs model in curved
space \cite{W},
\begin{equation}
S=4\pi\int dt\int^\infty_0dr\left[-\frac{1}
{4}r^2f_{\mu\nu}f^{\mu\nu}-(D_\mu \chi)^*D^\mu \chi
-\frac{g^2}{2r^2}\left(|\chi |^2-\frac{1}{g^2}\right)^2\right]\ ,
\label{2.3}
\end{equation}
where
\begin{equation}
f_{\mu\nu}=\partial_\mu a_\nu-\partial_\nu a_\mu\ ,
\label{2.4}
\end{equation}
is the field strength,
\begin{equation}
\chi=\alpha+i\Big(\gamma-\frac{1}{g}\Big)\ ,
\label{2.5}
\end{equation}
is a complex scalar and
\begin{equation}
D_\mu \chi=(\partial_\mu-iga_\mu)\chi\ ,
\label{2.6}
\end{equation}
is the covariant derivative.  The indices are raised and lowered with the
$1+1$ dimensional metric $ds^2 = -dt^2+dr^2$.  (Note that in this paper we
raise and
lower indices with a flat space metric; we do not absorb the explicit factors
of $r$ into a curved space metric as was done in Ref. \cite{FKS}.)

The equations of motion for the reduced theory are
\begin{mathletters}
\begin{equation}
-\partial^\mu(r^2f_{\mu\nu})=ig\left[(D_\nu \chi)^*\chi-\chi^*D_\nu
\chi\right]\ ,
\label{2.7a}
\end{equation}
and
\begin{equation}
\left[-D^2+\frac{1}{r^2}(g^2|\chi|^2-1)\right]\chi=0\ .
\label{2.7b}
\end{equation}
\end{mathletters}
It is convenient to write the complex field $\chi$ in polar form
\begin{equation}
\chi(r,t)=-i\rho(r,t)\exp[ig\varphi(r,t)]\ ,
\label{2.8}
\end{equation}
which casts the equations of motion in terms of $\rho\, ,\,\varphi$ and
$a_\mu$ as
\begin{mathletters}
\begin{equation}
\partial^\mu(r^2f_{\mu\nu})+2g^2\rho^2(\partial_\nu \varphi-a_\nu)=0\ ,
\label{2.9a}
\end{equation}
\begin{equation}
\partial_\mu \partial^\mu \rho-g^2 \rho(\partial^\mu \varphi-a^\mu)
(\partial_\mu \varphi-a_\mu)-\frac{1}{r^2}(g^2 \rho^2 -1)\rho=0\ ,
\label{2.9b}
\end{equation}
and
\begin{equation}
\partial^\mu \left[\rho^2 (\partial_\mu \varphi-a_\mu)\right]=0\ .
\label{2.9c}
\end{equation}
\end{mathletters}
Note that (\ref{2.9c}) follows from (\ref{2.9a}) so there are three, not four,
independent equations of motion which is expected because of the residual
$U(1)$ gauge invariance.

In $1+1$ dimensions $f_{\mu\nu}$ must be proportional to $\epsilon_{\mu\nu}$
so we can define the gauge invariant field $\psi$ through
\begin{equation}
r^2f_{\mu\nu}=-2\epsilon_{\mu\nu} \psi \ .
\label{2.10}
\end{equation}
(Here $\epsilon_{01}=+1$.)  Equation (\ref{2.9a}) now becomes
\begin{equation}
\partial^\alpha \psi=-g^2\epsilon^{\alpha\nu} \rho^2
(\partial_\nu \varphi-a_\nu)\ ,
\label{2.11}
\end{equation}
which implies
\begin{mathletters}
\begin{equation}
\partial_\mu \left(\frac{\partial^\mu \psi}{\rho^2}\right)-\frac{2g^2}
{r^2}\psi=0\ .
\label{2.12a}
\end{equation}
Substituting (\ref{2.11}) into (\ref{2.9b}) gives
\begin{equation}
\partial_\mu \partial^\mu \rho+\frac{1}
{g^2\rho^3}\partial_\mu \psi \, \partial^\mu \psi-\frac{1}{r^2}
(g^2\rho^2-1)\rho=0\ .
\label{2.12b}
\end{equation}
\end{mathletters}

Equations (\ref{2.12a}) and (\ref{2.12b}) are the equations of motion for the
gauge invariant variables $\rho$ and $\psi$.  If a gauge is chosen then the
gauge-variant variables $\varphi$ and $a_\mu$ can be determined using
(\ref{2.11}).  For example, in the $\varphi=0$ gauge, given $\rho$ and
$\psi$, equation (\ref{2.11}) directly determines $a_\nu$.  Equations
(\ref{2.12a}) and (\ref{2.12b}) for $\psi$ and $\rho$ can be solved after
specifying initial value data, that is the values of $\psi\, ,\,\dot\psi\,
,\,\rho$ and $\dot\rho$ at some fixed time.  Any initial value data
expressed in terms of $\psi\, ,\,\dot\psi\, ,\,\rho$ and $\dot\rho$ will
be consistent with the Gauss's law constraint.  This is because Gauss's law is
the $\nu=0$ component of equation (\ref{2.9a}) which is equivalent to the
$\alpha=1$ component of (\ref{2.11}).

Using the equations of motion, the energy associated with the action
(\ref{2.2}) can be written in terms of $\rho$ and $\psi$ as
\begin{equation}
E=8\pi\int^\infty_0 dr \Bigg[ \frac{1}{2}(\partial_t \rho)^2 +\frac{1}{2}
(\partial_r \rho)^2+\frac{1}{2g^2 \rho^2}(\partial_t \psi)^2 +\frac{1}
{2g^2 \rho^2}(\partial_r \psi)^2
+\frac{\psi^2}{r^2}+\frac{(g^2 \rho^2 -1)^2}{4g^2 r^2}\Bigg]\ .
\label{2.13}
\end{equation}
{}From this expression
we see that the field configuration with zero energy has $\psi=0$ and
$\rho=1/g$.

It is convenient to introduce a shifted field $\delta$
defined by
\begin{equation}
2g\delta=1-g^2 \rho^2 \ .
\label{2.14}
\end{equation}
Equations (2.12) now become
\begin{mathletters}
\label{2.15}
\begin{equation}
\partial_\mu\partial^\mu\psi-\frac{2}{r^2}\,\psi=-2g
\partial_\mu \left(\frac{\delta\, \partial^\mu \psi}{1-2g\delta}\right)
\label{2.15a}
\end{equation}
and
\begin{equation}
\partial_\mu \partial^\mu \delta-\frac{2}{r^2}\,\delta=
g\frac{\partial_\mu\psi\, \partial^\mu\psi}{1-2g
\delta}
-g\frac{\partial_\mu\delta\,
\partial^\mu\delta}{1-2g\delta}
-\frac{4g}{r^2}\,\delta^2\ ,
\label{2.15b}
\end{equation}
\end{mathletters}
and the energy (\ref{2.13}) is
\begin{equation}
E=8\pi\int_0^\infty
dr\Bigg[\frac{1}{2(1-2g\delta)}
\left\{(\partial_t \delta)^2+(\partial_r \delta)^2
+(\partial_t \psi)^2+(\partial_r \psi)^2 \right\}
+\frac{\psi^2}{r^2}+\frac{\delta^2}{r^2}\Bigg]\ .
\label{2.16}
\end{equation}

We will show how to solve (\ref{2.15}) in a perturbative expansion in $g$ after
we
show that these equations generically describe imploding shells in the far
past which turn into outgoing shells in the far future.
\vskip .3in

\section{Linearization in the Far Past and Far Future}
\label{sec3}

In the introduction we pointed out that in the spherical ansatz a typical
solution to the equations of motion describes an imploding shell, moving close
to the speed of light, at very early times.  To see how this comes about
consider equations (\ref{2.15}) and imagine that at some early time $T<0$,
$\delta$
and $\psi$ are both pulses of width $\Delta$ centered at $r$ near $|T|$ with
$\Delta \ll |T|$.  By a pulse we mean a function which is very close to zero
except in a region whose size is $\Delta$.  Accordingly, space and time
derivatives of $\psi$ and $\delta$ are pulses with amplitude of order
$\psi/\Delta$ and $\delta/\Delta$.
 For $r\simeq|T|\gg\Delta$ we can now neglect the $1/r^2$ terms in
(\ref{2.15}).  We
then see that if $\psi(r,t)$ and $\delta(r,t)$ depend only on $r+t$, that is,
$\psi(r,t)=\psi_p(r+t)$ and $\delta(r,t)=\delta_p(r+t)$
then equations (\ref{2.15}) are
satisfied.  If $\psi_p(u)$ and $\delta_p(u)$ are close to zero
except for $u$ in a
region of size $\Delta$ around $u=0$, then the solution describes an
incoming shell of width $\Delta$ moving undistorted along $r=-t$.  This
description remains valid for all $t \ll -\Delta$.

Although equations (\ref{2.15}) with the $1/r^2$ terms neglected have
solutions
$\psi(r,t)=\psi_p(r+t)$ and $\delta(r,t)=\delta_p(r+t)$,
these equations are not  linear.
We are for example not assuming that $\delta$ and $\psi$ are
small.  There is however a sense in which we can say that the theory
linearizes.  Consider the field strength $F_{\mu\nu}$ in the $3+1$ dimensional
theory given by (\ref{1.3}).  We now argue that there is a choice of gauge
such that the commutator term, $[A_\mu,A_\nu]$,
can be neglected relative to the linear
term, $\partial_\mu A_\nu-\partial_\nu A_\mu$, for all $t$ much less than
$-\Delta$.  Clearly this statement requires a gauge choice since the size of
$A_\mu$ and its derivatives are gauge dependent.  Nonetheless the statement
that there exists a gauge with this property is a gauge {\it invariant\/}
characterization of solutions in the spherical ansatz.

Work in the $A_0=0$ gauge which in the reduced theory is equivalent to the
$a_0=0$ gauge.  Using (\ref{2.11}) and (\ref{2.14}) we then have
\begin{equation}
\partial_r \psi(r,t)
=
\left( 2 g \delta(r,t)-1 \right) \partial_t \varphi(r,t) ~~,
\label{3.1}
\end{equation}
or
\begin{equation}
\varphi(r,t)= -\int_{-\infty}^t
dt'\,\frac{\partial_r \psi(r,t')}{1- 2 g \delta (r,t')  }
+ \varphi_0 (r)
\label{3.2}
\end{equation}
where $\varphi_0(r)$ is an arbitrary function of $r$.  Now for $t\ll-\Delta$,
using $\psi_p(r+t)$ and $\delta_p(r+t)$ for $\psi$ and $\delta$, we have
\begin{equation}
\varphi(r,t)= -\int_{-\infty}^{r+t} du \,
\frac{\partial_u \psi_p (u)}{1- 2 g \delta_p (u)}
+ \varphi_0 (r).
\label{3.3}
\end{equation}
Equations (\ref{2.11}) and (\ref{2.14}) also give, for all time,
\begin{equation}
a_1(r,t)=\partial_r \varphi(r,t)+
\frac{\partial_t \psi(r,t)}{ 1 - 2 g \delta (r,t) } ~~,
\label{3.4}
\end{equation}
which for $t\ll-\Delta$, upon using (\ref{3.3}), becomes
\begin{equation}
a_1(r,t)=\partial_r \varphi_0 (r)\ .
\label{3.5}
\end{equation}
In the $a_0=0$ gauge we are still free to make a
time independent gauge transformation which we do to set
\begin{equation}
a_1 (r,t)=0
\label{3.6}
\end{equation}
for $t\ll-\Delta$.    What we have shown is that in the $a_0=0$ gauge if we
set $a_1=0$ at one early time then we have $a_1=0$ at all early times.

If we now turn to (\ref{2.1}) we see that in the gauge we have chosen,
at early
times, the dominant pieces in $A_i$ go like $\alpha/r$ and $\gamma/r$.  Since
the gauge has been specified, $\alpha$ and $\gamma$ are determined by $\delta$
and $\psi$ so, for example, derivatives of $\alpha$ and $\gamma$ are of order
$\alpha/\Delta$ and $\gamma/\Delta$.  This means that $\partial_\mu A_i$ is of
order $\alpha/(r\Delta)$ and $\gamma/(r\Delta)$ whereas $A_iA_j$ is of order
$\alpha^2/r^2$, $\alpha\gamma/r^2$ and $\gamma^2/r^2$.  For $\Delta\ll r$
the commutator term in $F_{\mu\nu}$ can be neglected relative to the
linear term.  (In fact $F_{\mu\nu}$ goes like $1/(r\Delta)$ so the total
energy in the shell scales as $1/\Delta$.)  We also see that the energy is,
at early
times, quadratic in the field $A_i$ which is a signature of a linear
theory.

We are particularly interested in solutions which in the far past describe
free particles.  Thus we want the gauge field $A_\mu$, in the gauge in which
the solution linearizes in the far past, to be independent of $g$ in the far
past.   In this case $F_{\mu\nu}$ is also
independent of $g$ since
$F_{\mu\nu} \simeq \partial_\mu A_\nu - \partial_\nu A_\mu$.
Then, for example, the energy which depends on $F_{\mu\nu}^{\,2}$ is
independent of $g$.

In the gauge in which the solutions linearize, $a_0=0$ and $a_1 = 0$ for
$t \ll -\Delta$.  For $A_\mu$ in this regime to be independent of $g$, we
require that $\alpha$ and $\gamma$ be independent of $g$ (see (\ref{2.1})).
Now from (\ref{2.5}), (\ref{2.8}) and (\ref{2.14}), we have
\begin{equation}
\delta = \gamma - \frac{1}{2} g (\alpha^2 + \gamma^2) ~~.
\label{3.7}
\end{equation}
Using (\ref{2.5}), (\ref{2.8}), (\ref{2.14}) and (\ref{3.1}), we also have
\begin{equation}
\partial_r \psi
= -\partial_t \alpha
+ g (\gamma \partial_t \alpha - \alpha \partial_t \gamma) ~~.
\label{3.8}
\end{equation}
This means that if in the far past $\alpha$ and $\gamma$ are independent of
$g$, then the gauge invariant variables $\psi$ and $\delta$ have pieces which
are zeroth and first order in $g$.  If $\alpha(r,t)$ and $\gamma(r,t)$ for
$t \ll -\Delta$ are described by pulses $\alpha(r,t) = \alpha_p(r+t)$
and $\gamma(r,t) = \gamma_p(r+t)$ then
\begin{mathletters}
\label{3.9}
\begin{equation}
\delta_p(u) = \gamma_p(u) - \frac{1}{2} g
\left( \alpha_p^{\,2} (u) + \gamma_p^{\,2}(u) \right)
\label{3.9a}
\end{equation}
and
\begin{equation}
\partial_u \, \psi_p(u) = -\partial_u \, \alpha_p(u)
+ g \Bigl(
\gamma_p (u) \partial_u \alpha_p(u) -
\alpha_p (u) \partial_u \gamma_p(u)
\Bigr) \ .
\label{3.9b}
\end{equation}
\end{mathletters}
Note that if we assume this form for $\delta_p(u)$ and $\psi_p(u)$, then the
energy evaluated using (\ref{2.16}), neglecting the $1/r^2$ terms, is
indeed independent of $g$.

We have shown that an incoming pulse moving undistorted at the speed of light,
$\psi (r,t) = \psi_p(r+t)$ and $\delta(r,t) = \delta_p(r+t)$,
solves the equation of motion (\ref{2.15}), as long as the pulse is located
at a radius much larger than the width of the pulse.  Similarly an outgoing
pulse, $\psi(r,t) = \psi_f(t-r)$ and $\delta(r,t) = \delta_f(t-r)$ solves the
equations of motion under the same conditions.  A general scattering solution
consists of an incoming pulse in the far past which in the far future is an
outgoing pulse.  The future pulse profiles, $\psi_f$ and $\delta_f$ will not,
in general, coincide with the past pulse profiles $\psi_p$ and $\delta_p$.  In
fact a full understanding of the solutions to (\ref{2.15}) would allow us to
determine $\psi_f$ and $\delta_f$ given arbitrary $\psi_p$ and $\delta_p$.

\vskip .3in

\section{The Perturbative Expansion}
\label{sec4}

In this section we solve equations (\ref{2.15})
perturbatively in $g$ in an in-field formalism
 where we specify the form of
the solution in the far past.
If we set $g$ equal to zero, equations (\ref{2.15a}) and (\ref{2.15b}) have the
same form:
\begin{mathletters}
\label{4.1}
\begin{equation}
\left(-\partial^2_t +\partial^2_r -\frac{2}{r^2}\right)\psi_0 =0 ,
\label{4.1a}
\end{equation}
and
\begin{equation}
\left(-\partial^2_t +\partial^2_r -\frac{2}{r^2}\right)\delta_0 =0\ .
\label{4.1b}
\end{equation}
\end{mathletters}
It is convenient to switch to light cone coordinates defined as
\begin{eqnarray}
u &=& t+r\nonumber\\
v &=& t-r \ ,\label{4.2}\end{eqnarray}
in which $ds^2 = -dt^2+dr^2 = -du~dv$ and $\partial_\mu \partial^\mu =
-\partial_t^2 + \partial_r^2 = -4\partial_u \partial_v$.
Finite energy requires the solutions to vanish at $r=(u-v)/2 =0$.  The
general finite energy
solution to the $g=0$ homogeneous equations (\ref{4.1}) is
\begin{mathletters}
\label{4.3}
\begin{equation}
\psi_0 (u,v)=G'(u)+G'(v)-\frac{2}{u-v}\left[G(u)-G(v)\right]
\label{4.3a}
\end{equation}
and
\begin{equation}
\delta_0 (u,v)=H'(u)+H'(v)-\frac{2}{u-v}\left[H(u)-H(v)\right] \ ,
\label{4.3b}
\end{equation}
\end{mathletters}
where $G(x)$ and $H(x)$ are two arbitrary functions
which approach constants as $x$ goes to $\pm\infty$.
If $G'(x)$ and $H'(x)$ are both nonzero
only in a region of size $\Delta$ containing $x=0$,
then the solution represents an energy shell of width $\Delta$ moving
along $r=-t$ at early times ($t\ll-\Delta$) and along $r=t$ at late times
($t\gg\Delta$).

We now expand the fields $\psi$ and $\delta$ in the coupling $g$:
\begin{mathletters}
\begin{eqnarray}
\psi &=& \psi_0 +g\psi_1 +g^2 \psi_2 +\ldots\ ,
\label{4.4a}\\
\delta &=& \delta_0 +g\delta_1 +g^2 \delta_2 +\ldots\ ,
\label{4.4b}
\end{eqnarray}
\end{mathletters}
and substitute into the equations of motion (\ref{2.15}).
Expanding the equations in $g$ gives
\begin{eqnarray}
\left[ \partial_u \partial_v + \frac{2}{(u-v)^2} \right]\psi_1 &=&
I^\psi_1\ ,
\nonumber\\
\left[ \partial_u \partial_v + \frac{2}{(u-v)^2} \right]\delta_1\, &=&
I^\delta_1\ ,
\nonumber\\
 &\vdots&
\label{4.5}\\
\left[ \partial_u \partial_v + \frac{2}{(u-v)^2} \right]\psi_n &=& I^\psi_n\ ,
\nonumber\\
\left[ \partial_u \partial_v + \frac{2}{(u-v)^2} \right]\delta_n \,
&=& I^\delta_n\ ,
\nonumber
\end{eqnarray}
where $I^\psi_n$ and $I^\delta_n$ both
depend only on $\psi_0,\ldots\psi_{n-1}$ and
$\delta_0,\ldots\delta_{n-1}$.  For example
\baselineskip=30pt
\begin{mathletters}
\begin{eqnarray}
I_1^\psi &=& \frac{1}{2}\partial_\mu (\delta_0\, \partial^\mu \psi_0 )
\label{4.6a}\\
I_1^\delta &=& \frac{1}{4}
\partial_\mu \delta_0\, \partial^\mu \delta_0 -\frac{1}{4}\partial_\mu \psi_0
\, \partial^\mu
\psi_0 +\frac{4\delta_0^2}{(u-v)^2}
\label{4.6b}\\
I_2^\psi &=& \frac{1}{2}\partial_\mu ( \delta_1 \, \partial^\mu \psi_0 +
\delta_0 \, \partial^\mu \psi_1  +2\delta_0^2 \, \partial^\mu \psi_0 )
\label{4.6c}\\
I_2^\delta &=&
\frac{1}{2}\delta_0\,\partial_\mu \delta_0 \, \partial^\mu
\delta_0
-\frac{1}{2}\delta_0 \,\partial_\mu \psi_0 \, \partial^\mu \psi_0
-\frac{1}{2}\partial_\mu \psi_1 \,
\partial^\mu \psi_0
+\frac{1}{2}\partial_\mu \delta_1 \, \partial^\mu
\delta_0 +\frac{8\delta_1 \delta_0}{(u-v)^2}\ .
\label{4.6d}
\end{eqnarray}
\end{mathletters}
\baselineskip=24pt plus 2pt minus 2pt

Given the sources $I_n^\psi$ and $I_n^\delta$, equations (\ref{4.5}) do not
uniquely specify $\psi_n$ and $\delta_n$ because we can add homogeneous
solutions of the form (\ref{4.3a}) and (\ref{4.3b}) to $\psi_n$ and $\delta_n$
and still satisfy the equations.  We will first solve (\ref{4.5}) with the
requirement that the solutions vanish in the far past,
and will later use the freedom to add homogeneous solutions
to find solutions which describe $g$-independent $\alpha$
and $\gamma$ pulses
in the far past.  Consider the equation
\begin{equation}
\left[ \partial_u \partial_v + \frac{2}{(u-v)^2} \right]\Phi(u,v)=I(u,v)
\label{4.7}\end{equation}
where the source $I\to 0$ as $v\to -\infty$,  which we solve requiring
that the solution $\Phi\to 0$ as $v\to -\infty$,
and also that
$\Phi$ vanishes at $r=(u-v)/2=0$.  The retarded
Green's function solution is
\begin{equation}
\Phi (u,v)=
\int_v^u du'\int_{-\infty}^v dv'\,F(u,v;u',v')I(u',v')
\label{4.8}\end{equation}
where
\begin{equation}
F(u,v;u',v')=1-\frac{2(u-u')(v-v')}{(u-v)(u'-v')}\ .
\label{4.9}\end{equation}
(Note that $F(u,v;u',v')$ satisfies (\ref{4.7}) with $I=0$.)

We can use (\ref{4.8}) to get explicit formulas for $\psi_1$ and $\delta_1$
given the sources $I_1^\psi$ and $I_1^\delta$ of (\ref{4.6a}) and
(\ref{4.6b}) along with $\psi_0$ and $\delta_0$ expressed in terms of the
arbitrary functions $G(x)$ and $H(x)$ of (\ref{4.3a}) and (\ref{4.3b}).  The
result is
\begin{mathletters}
\label{4.10}
\begin{eqnarray}
\psi_1(u,v) &=&
-G'(u)H'(v)-G'(v)H'(u)-\frac{6}{(u-v)^2}[G(u)-G(v)][H(u)-H(v)]\nonumber\\
& & +\frac{2}{u-v}[G'(u)+G'(v)][H(u)-H(v)]\nonumber\\
& & +\frac{2}{u-v}[H'(u)+H'(v)][G(u)-G(v)]
\label{4.10a}\end{eqnarray}
and
\begin{eqnarray}
\delta_1(u,v) &=&
G'(u)G'(v)+\frac{[G(u)-G(v)]^2}{(u-v)^2}-\frac{2}{u-v}[G'(u)
+G'(v)][G(u)-G(v)]\nonumber\\
& & +\frac{2}{u-v}\int_v^u
dx\,G'(x)^2-H'(u)H'(v)-\frac{5}{(u-v)^2}[H(u)-H(v)]^2\nonumber\\
& & +\frac{2}{u-v}[H'(u)+H'(v)][H(u)-H(v)]+\frac{2}{u-v}\int_v^u
dx\,H'(x)^2\ .
\label{4.10b}\end{eqnarray}
\end{mathletters}
Using the fact that $G'(x)$ and $H'(x)$ are only nonvanishing near $x=0$ we
see that $\psi_1$ and $\delta_1$ do in fact go to zero as $v\to-\infty$.
We
are still free to add homogeneous pieces to $\psi_1$ and $\delta_1$ if we so
choose.  Indeed we will need to use this freedom since
equations (\ref{3.9})
and the requirement that $\alpha$ and $\gamma$ are $g$-independent in
the far past imply  that $\psi_0$,
$\psi_1$, $\delta_0$, and $\delta_1$ are in general nonzero
as $v\to -\infty$.
Given $\psi_1$ and $\delta_1$ we can evaluate $I_2^\psi$ and
$I_2^\delta$ in (\ref{4.6c}) and (\ref{4.6d}) and then obtain $\psi_2$ and
$\delta_2$ using (\ref{4.8}).  Continuing along these lines we can, in
principle, get $\psi$ and $\delta$ to any order in $g$.  Of course in practice
the calculations get more complicated with each order of perturbation
theory.

\vskip .3in

\section{Topological Charge}
\label{sec5}

In the spherical ansatz, using the equations of motion, the local topological
charge (\ref{1.1}) can be written \cite{FKS} as
\begin{equation}
Q(R)=\frac{g}{2\pi}\int^\infty_{-\infty}dt\int^R_0 dr\,\left(-\partial^2_t+
\partial^2_r-\frac{2}{r^2}\right)\psi \ ,
\label{5.1}
\end{equation}
where $\psi(r,t)$ is a solution to (\ref{2.15}).
Since the integrand in (\ref{1.1}) is a total divergence we can also write
$Q(R)$ as a surface integral.  Using (\ref{2.15a}) we have that
\begin{equation}
Q(R) = \frac{-g^2}{\pi}
\int\limits_{-\infty}^{\infty} dt
\int\limits_0^R dr ~ \partial_\mu
\left\{
\frac{\delta\, \partial^\mu \psi}{1 - 2 g \delta}
\right\}
{}~,
\label{5.2}
\end{equation}
or
\begin{equation}
Q(R) = \frac{-g^2}{\pi}
\int\limits_{-\infty}^{\infty} dt ~
\frac{\delta\, \partial_r \psi}{1 - 2 g \delta}  \Biggr|_{r=R} ~~,
\label{5.3}
\end{equation}
since the contributions from the $t=\pm\infty$ surfaces
and the $r=0$ surface vanish.

We are interested in $R$ much larger than any length scale associated with the
solution.  At this large value of $R$, the integral in (\ref{5.3}) is only
nonzero
when the incoming pulse passes $R$ at $t \simeq -R$ and when the outgoing
pulse passes $R$ at $t\simeq R$.
For
$t \simeq -R$
we have
$\psi(R,t) = \psi_p(R+t) + {\cal O}(1/R)$
and
$\delta(R,t) = \delta_p(R+t) + {\cal O}(1/R)$
while for
$t \simeq R$
we have
$\psi(R,t) = \psi_f(t-R) + {\cal O}(1/R)$
and
$\delta(R,t) = \delta_f(t-R) + {\cal O}(1/R)$.
Here $\psi_p (u) = \psi(u,-\infty)$ and
$\delta_p (u)=\delta(u,-\infty)$ are the incoming pulse functions described in
section \ref{sec3} and $\psi_f (v) = \psi (\infty ,v) $ and $\delta_f (v) =
\delta (\infty ,v)$ are the associated future pulse
functions.  Substituting into (\ref{5.3}) and taking the limit as $R$ goes to
infinity gives
\begin{equation}
Q = \frac{g^2}{\pi} \int\limits_{-\infty}^{\infty} dv
\left\{
\frac{\delta_f (v)\,\partial_v \psi_f (v) }{1 - 2 g \delta_f (v)}
\right\} ~-~\frac{g^2}{\pi} \int\limits_{-\infty}^{\infty} du
\left\{\frac{\delta_p (v)\,\partial_u \psi_p(u) }{1
- 2 g \delta_p (u)}
\right\} ~~.
\label{5.4}
\end{equation}
Note that topological charge develops only to the extent that the
outgoing profiles differ from the incoming profiles.

We can develop (\ref{5.4}) perturbatively in $g$ if we use the perturbative
expansion for the solutions $\psi(u,v)$ and $\delta(u,v)$.  To zeroth order in
$g$, $\psi(u,v)$  and $\delta(u,v)$ are given by (\ref{4.3a}) and (\ref{4.3b})
so we see that to this order, the future and past profile functions agree,
that is,
$\psi_{f0}(x) = \psi_{p0}(x)$
and
$\delta_{f0}(x) = \delta_{p0}(x)$.
This means that the topological charge vanishes to second order in $g$.

If we look at (\ref{1.2}) along with (\ref{1.3}) and (\ref{1.4}) we see that
the second order topological charge, outside the spherical ansatz, can be
written as
\begin{equation}
Q^{[2]} = \frac{g^2}{16\pi^2} \sum\limits_{a=1}^{3} \int d^4x ~
\epsilon^{\mu\nu\alpha\beta} \partial_\mu A_\nu^{\,a}
\partial_\alpha A_\beta^{\,a} \ ,
\label{5.5}
\end{equation}
where
\begin{equation}
\partial^\mu (\partial_\mu A_\nu^{\,a} - \partial_\nu A_\mu^{\,a}) = 0\ .
\label{5.6}
\end{equation}
Thus the second order non-Abelian topological charge is equivalent to the
topological charge in an Abelian theory.  In appendix A we show, without
restricting to the spherical ansatz, that for localized energy pulses which
solve Maxwell's equations, the topological charge vanishes.

To get the
third order topological charge we need $\psi_1(u,v)$ and $\delta_1(u,v)$ whose
general form is given by (\ref{4.10}) plus possible homogeneous pieces.
With
$\psi_1$ and $\delta_1$
given by (\ref{4.10}), we see that
both go to zero for
large positive $u$
and for large negative $v$
so that
$\psi_{f1} = \psi_{p1} =  \delta_{f1} = \delta_{p1} = 0$.
Any homogeneous piece in $\psi_1$ and $\delta_1$ will make cancelling
contributions to (\ref{5.4}) so we have that to order $g^3$ the topological
charge vanishes within the spherical ansatz.

The fourth order topological charge is more difficult to evaluate.  For
arbitrary initial profiles we would need to find $\psi_2$ and $\delta_2$ which
entails evaluating (\ref{4.8}) with the sources
(\ref{4.6c}) and (\ref{4.6d}) where $\psi_0$ and $\delta_0$ are given by
(\ref{4.3}) and $\psi_1$ and $\delta_1$ are given by (\ref{4.10})
plus possible homogeneous terms.  We have
done this only in examples.  For
the examples considered we find that
$\psi_{f2}$ and $\delta_{f2}$ are non-vanishing even though
$\psi_{p2} = \delta_{p2} = 0$.
Nonetheless, in every case
the fourth order topological charge vanishes because of
cancellations in the first term in (\ref{5.4}).
Our goal is not to prove that the fourth order topological
charge is zero for arbitrary initial profiles.
Our goal is to show that there are initial profiles
for which $Q$ is of order $g^n$ for some finite $n$.
Therefore, onward to the $g^5$ term.

We have evaluated the fifth order topological charge
for several initial profiles. In every
case, we have found a nonzero result.
In principle,
once the initial conditions are specified
the calculation of the fifth order
topological charge follows from (\ref{5.4}) and the
results of section IV.
In practice,
at this order the calculations are rather involved and
we relegate the details to an appendix.
Here, we will describe the initial conditions
we use, and quote the resulting $Q^{[5]}$ for two examples.

Initial conditions are specified by the functions
$\gamma_p(u)$ and $\alpha_p(u)$ of equations (\ref{3.9})
which we require to be $g$-independent.
We make the choice
$\gamma_p(u)=0$ because it simplifies the calculation
considerably. From  (\ref{3.9}) and (\ref{4.3}),
this choice implies $G'(x)=-\alpha_p(x)$ and $H(x)=0$.
Because $H(x)=0$, equation (\ref{4.10a}) gives $\psi_1(u,v)=0$,
which agrees with the order $g$
term of (\ref{3.9b}).
In general,
the equations of
motion (\ref{2.15})
are invariant under the transformation
$g \to -g$ and $\delta \to -\delta$ with $\psi$ fixed.
For $\gamma_p=0$, the initial conditions (\ref{3.9})
are invariant also.
This means that
for $\gamma_p=0$, the expansion in $g$ of $\delta$ can
contain only odd powers of $g$, while that of $\psi$
can contain only even powers of $g$.
The order $g$ term of (\ref{3.9a}) requires
that $\delta_{p1}(u)=-\frac{1}{2}G'(u)^2$.
Therefore, we must add the homogeneous term
\begin{equation}
\delta_1^{{\rm hom}}(u,v)=-\frac{1}{2}\left[ G'(u)^2 + G'(v)^2
-\frac{2}{u-v}\int_v^u
dx\,G'(x)^2 \right]
\label{5.7}
\end{equation}
to $\delta_1$ of  (\ref{4.10b}) to obtain
\begin{eqnarray}
\delta_1(u,v) &=&
-\frac{1}{2}[G'(u)-G'(v)]^2
+\frac{[G(u)-G(v)]^2}{(u-v)^2} \nonumber\\
& & -\frac{2}{u-v}[G'(u)+G'(v)][G(u)-G(v)]
+  \frac{3}{u-v}\int_v^u dx\,G'(x)^2
\ . \label{5.8}
\end{eqnarray}
Thus, choosing $\gamma_p=0$ means that $\delta_0$ and
$\psi_1$ are zero for all $u$ and $v$, while $\psi_0(u,v)$
and $\delta_1(u,v)$ are given in terms of the single
function $G(x)$ by equations (\ref{4.3a}) and (\ref{5.8}).

In appendix B, we show that for $\gamma_p=0$,
the fifth order term in the topological charge (\ref{5.4}) can
be written as the double integral
\begin{eqnarray}
Q^{[5]} &=& \frac{g^5}{\pi} \int_{-\infty}^\infty du dv~ \Biggl\{
 \frac{1}{2} \left[
  \frac{1}{2}
\partial_\mu\psi_0\, \partial^\mu\delta_1
 +\frac{4\psi_0\delta_1}{(u-v)^2}
\right]\Biggl[2 G'(u)G'(v) +
\frac{6}{(u-v)^2}[G(u)-G(v)]^2 \nonumber\\
& & \qquad\qquad\qquad\qquad\qquad\qquad\qquad\qquad\qquad
-\frac{4}{(u-v)^2}[G'(u)+G'(v)][G(u)-G(v)]\Biggr] \nonumber\\
& & \qquad\qquad\qquad
-\frac{3}{2} \frac{\psi_0 \delta_1}{(u-v)^2} - \frac{\psi_0^3 \delta_1}
 {(u-v)^2}
 - \frac{1}{8}\frac{\psi_0^5}{(u-v)^2} \Biggr\}\ .
\label{5.9}\end{eqnarray}
We have evaluated equation (\ref{5.9}) analytically for several
functions $G(x)$.  Here, we present the results for two examples.
For
\begin{equation}
G(x)=-\frac{x}{(1+x^2)}
\label{5.10}
\end{equation}
we find
\begin{equation}
Q^{[5]} = -\frac{15 \pi^2}{1024}~\frac{g^5}{\pi} \ .
\label{5.11}
\end{equation}
For
\begin{equation}
 G'(x)= \cases{0~;                 &  $x \leq -1$  \cr
                           (1-x^2)^2 ~;  & $-1 \leq x \leq1$  \cr
                          0~;                 & $1 \leq x $ \cr
                         }.
\label{5.12}
\end{equation}
we find
\begin{equation}
Q^{[5]} = \frac{1336082432}{11402015625}~\frac{g^5}{ \pi} \ .
\label{5.13}
\end{equation}

We have demonstrated that spherical solutions
to the Minkowski space classical equations of motion for pure $SU(2)$
which propagate freely in the far past can have nonzero
topological charge.  The power series expansion of $Q$ in $g$
can be nonzero at finite order  --- in the examples we
have presented, at order $g^5$.

\vskip .3in

\section{Fermion Production in a Background Field with\protect\\
Fractional Topological Charge}
\label{sec6}

The solutions which we consider in this paper have the property that the
topological charge inside a sphere of radius $R$, $Q(R)$ given by
(\ref{1.1}), approaches a fixed value as $R$ gets large  and the limiting
value,
$Q=\lim_{R\to\infty}Q(R)$, need not be an integer.  In fact we have shown that
$Q$ can be developed as a power series in $g$ so that $Q$ can be made small by
making $g$ small.  Suppose that we have a quantized left-handed fermi field,
$\hat\Psi$, coupled to this classical
non-Abelian background.  Without introducing a Higgs field,
we cannot give the fermion a gauge invariant mass, so
we take the fermion to be massless.
 Due to the anomaly,
the fermion number current $\hat J^\mu=
{}~ : {\hat{\bar\Psi}}\gamma^\mu\hat\Psi :$ is not
conserved, that is,
\begin{equation}
\partial_\mu \hat J^\mu =\frac{g^2}{32\pi^2}\, \epsilon^{\mu \nu\alpha \beta}
{\rm Tr}(F_{\mu \nu}F_{\alpha\beta})\ .
\label{6.1}\end{equation}
The question arises of how many fermions are produced in the presence of a
background field of the kind we have described.

Define the fermion number
operator, inside the sphere of
radius $R$, as
\begin{mathletters}
\begin{equation}
\hat N(R,t) = \int_0^R r^2\, dr \int d\cos\theta \, d\phi \,
\hat J^0(r,\theta,
\phi,t)
\label{6.2a}
\end{equation}
and the flux operator, through the two-sphere of radius $R$, as
\begin{equation}
\hat F(R,t) = \int d\cos\theta \, d\phi \, R^2 \,
\hat J^r(R,\theta,\phi,t)\ .
\label{6.2b}
\end{equation}
\end{mathletters}
(We will for the moment ignore the fact that
these are not well defined operators and return
to this point later.)
Now, integrating (\ref{6.1}) from $t=-T_0$ to $t=T_0$ and
over the inside of the sphere of radius $R$,
we obtain
\begin{equation}
\hat N(R,T_0)-\hat N(R,-T_0) + \int_{-T_0}^{T_0} dt\, \hat F(R,t)
= \frac{g^2}{32 \pi^2}\,
\epsilon^{\mu\nu\alpha\beta} \int_{-T_0}^{T_0} dt \int_{|\vec{x}|<R}
d^3x \, {\rm Tr}\left(F_{\mu\nu} F_{\alpha\beta}\right) \ .
\label{6.3}
\end{equation}
We are interested in background shells which at  early
times are moving along $r=-t$, enter the sphere of radius
$R$ at $t\simeq -R$, leave at $t\simeq R$, and then
continue to move outward along $r=t$.  If we take $T_0$
much larger than $R$, then the
right hand side of (\ref{6.3}) is well
approximated by $Q(R)$ of (\ref{5.1}) and we can write
\begin{equation}
\hat N(R,T_0)-\hat N(R,-T_0) + \int_{-T_0}^{T_0} dt\,
\hat F(R,t) = Q(R) \ .
\label{6.4}
\end{equation}

We are interested in tracing the evolution of a
normalized fermion state $|s\rangle$ in the background
of a shell which enters and leaves the sphere of radius
$R$.  Pick the state $|s\rangle$ such that at
$t=-T_0$ there are no fermions inside the sphere,
which implies
\begin{equation}
\langle s | \hat N(R,-T_0) | s \rangle =0 \ .
\label{6.5}
\end{equation}
The non-Abelian shell may have fermions
riding along with it which enter and leave the sphere
of radius $R$ making $\langle s | \hat F(R,t) | s \rangle$ nonzero
for $t\simeq -R$ and $t\simeq R$.  Any fermions anomalously
produced within the sphere will also contribute to
$\langle s | \hat F(R,t) | s \rangle$ when they leave the
sphere.  For $t\gg R$, the energy shell is far outside
the sphere and the background field is once again pure
gauge within the sphere.  Therefore, at late times
fermions inside the sphere are free, and
if we wait long enough they will leave the sphere.
Thus, for $T_0$ sufficiently large, we have
\begin{equation}
\langle s | \hat N(R,T_0) | s \rangle =0 \ .
\label{6.6}
\end{equation}
Taking the expectation of (\ref{6.4}) in the state $|s \rangle$, we obtain
\begin{equation}
\int_{-T_0}^{T_0} dt \,  \langle s | \hat F(R,t) | s \rangle = Q(R) \ .
\label{6.7}
\end{equation}

We interpret (\ref{6.7}) as follows.  Imagine a spherical detector
corresponding to the operator (\ref{6.2b}), which at any
time $t$ measures the flux of fermions passing out through
the two sphere of radius $R$.  (Fermions going in make a
negative contribution, as do anti-fermions going out.)
The detector is in continuous operation from
$t=-T_0$ to $t=T_0$ and therefore measures the total
fermion number passing through the two sphere.
For a given background field, we cannot predict
the outcome of an individual
measurement of the total fermion number passing through
the two sphere.  However, if we repeat
the measurement many times with the same background
field and the same state $|s\rangle$, then the
average of the observations will be $Q(R)$.

Although we believe that the spirit of the argument just
given is correct, it is specious
because the operators $\hat N(R,t)$ and $\hat F(R,t)$
have infinite vacuum fluctuations, that is,
\begin{mathletters}
\begin{equation}
\langle 0 | \hat N^2(R,t) | 0 \rangle = \infty
\label{6.8a}
\end{equation}
and
\begin{equation}
\langle 0 | \hat F^2(R,t) | 0 \rangle = \infty \ .
\label{6.8b}
\end{equation}
\end{mathletters}
Therefore, these operators also have infinite
fluctuations in any state in the Fock space; in
particular $\langle s | \hat N^2(R,t) | s \rangle$ is
infinite.  Thus, we cannot sensibly discuss the
measurement of these operators.

The infinite vacuum fluctuations arise
because of the sharp cutoffs in the $r$
integration at $r=R$
and in the $t$ integration at $t=\pm T_0$
in expressions (\ref{6.2a}), (\ref{6.2b}), and (\ref{6.3}).
This can be remedied by introducing
a smoothing function.  Let
\begin{equation}
f(r,t) = \cases{1, &if $0\leq r \leq R$ {\rm and} $|t|\leq T_0$ \cr
              0, &if $r \geq R_D$ {\rm or} $|t|\geq T_D$ \cr}
\label{6.9}
\end{equation}
and let $f(r,t)$ fall smoothly from $1$ to $0$
as $r$ varies from $R$ to $R_D$ and as $|t|$ varies
from $T_0$ to $T_D$.
Define the smoothed operators
\begin{mathletters}
\begin{equation}
\hat N_f = -\int_{-\infty}^{\infty} dt \int_0^\infty r^2\,
dr \int d\cos\theta \, d\phi \,
\frac{\partial f(r,t)}{\partial t} \hat J^0(r,\theta,
\phi,t)
\label{6.10a}
\end{equation}
and
\begin{equation}
\hat F_f = -\int_{-\infty}^{\infty} dt
\int_0^\infty r^2\, dr \int d\cos\theta \, d\phi \,
\frac{\partial f(r,t)}{\partial r} \hat J^r(r,\theta,\phi,t)\ .
\label{6.10b}
\end{equation}
\end{mathletters}
Because $\partial f(r,t) / \partial t$ is only nonzero for
$T_0\leq |t|\leq T_D$, it is also convenient to define
\begin{mathletters}
\begin{equation}
\hat N_f^{\rm \, early} = \int_{-T_D}^{-T_0} dt \int_0^\infty r^2\,
dr \int d\cos\theta \, d\phi \,
\frac{\partial f(r,t)}{\partial t} \hat J^0(r,\theta,
\phi,t)
\label{early}
\end{equation}
and
\begin{equation}
\hat N_f^{\rm \, late} = -\int_{T_0}^{T_D} dt \int_0^\infty r^2\,
dr \int d\cos\theta \, d\phi \,
\frac{\partial f(r,t)}{\partial t} \hat J^0(r,\theta,
\phi,t)
\label{late}
\end{equation}
\end{mathletters}
where $\hat N_f^{\rm \,late} - \hat N_f^{\rm \,early} = \hat N_f$.
Note that the eigenvalues
of $\hat N_f$, $\hat F_f$,
$\hat N_f^{\rm \,early}$, and $\hat N_f^{\rm \,late}$,
are not integers.
A short calculation
shows that the vacuum fluctuations of the smoothed currents
are given by
\begin{equation}
\int d^4 x \int d^4 y\, f(x) f(y) \langle 0 | \hat J^\mu (x) \hat J^\nu (y)
| 0 \rangle
= \frac{1}{3}\int \frac{d^4 k}{(2\pi)^5}\,\theta(k^0)\,\theta(k^2)\,
|\tilde f (k) |^2 \left( k^\mu k^\nu -
g^{\mu\nu}k^2 \right) \ ,
\label{fluctuation}
\end{equation}
where $\tilde f$ is the
fourier transform of $f$, and the
currents are understood to be normal ordered.
{}From (\ref{fluctuation}), we see that
by a judicious choice of $f(r,t)$,
the vacuum fluctuations
can be made arbitrarily small.  For the massless
fermion case which we are considering this
requires $R_D$ to be much larger than $R$
and $T_D$ to be much larger than $T_0$.
We also require that
$T_0\gg R_D$ so that during the measurements of $\hat N_f^{\rm \,early}$
and $\hat N_f^{\rm \, late}$
the energy shell is far outside the detector.
To summarize the relations between the relevant
length scales, we have $\Delta \ll R \ll R_D \ll T_0 \ll T_D$
where again $\Delta$ is the length which characterizes the pulse.

If we multiply both sides of (\ref{6.1}) by $f(r,t)$ and integrate,
we get
\begin{equation}
\hat N_f^{\rm \,late}- \hat N_f^{\rm \,early} + \hat F_f
= \frac{g^2}{32 \pi^2}\,
\epsilon^{\mu\nu\alpha\beta} \int_{-\infty}^{\infty} dt \int
d^3x\, f(r,t)\, {\rm Tr}\left( F_{\mu\nu} F_{\alpha\beta}\right) \ .
\label{6.11}
\end{equation}
Using the equations of motion for the fields in the
spherical ansatz, we can write the right hand side of
(\ref{6.11}) as
\begin{equation}
\frac{g}{2\pi} \int_{-\infty}^{\infty} dt \int_0^{\infty} dr \,
f(r,t) \left(-\partial^2_t+
\partial^2_r-\frac{2}{r^2} \right)\psi \ .
\label{6.12}
\end{equation}
For $r\gg\Delta$, and in particular
for $r\geq R$, the integrand of (\ref{6.12}) is well approximated by
$f(r,t)\left(-\partial^2_t+
\partial^2_r\right)\psi$ which is near zero because wherever and
whenever
$\psi$ is nonzero at large $r$, it
is given either by $\psi_p(t+r)$ or by $\psi_f(t-r)$ up to
terms of order $1/R$.
Also, for $|t|\geq T_0$ the shell is outside $r=R$ and so at
these early and late times
$\psi$ is close to zero for $r\leq R$.
Thus, the integrand of (\ref{6.12}) is close to zero wherever and whenever
$f(r,t)\neq 1$.  Therefore,
the expression (\ref{6.12}) is $Q$ up to terms of order $1/R$.

Taking the expectation of (\ref{6.11}) in a state $| s \rangle$
which has
\begin{equation}
\langle s | \hat N_f^{\rm \,early} | s \rangle =0
\label{6.15}
\end{equation}
and neglecting terms of order $1/R$, we obtain
\begin{equation}
\langle s | \hat F_f | s \rangle = Q \ ,
\label{6.16}
\end{equation}
where again we argue that
$\langle s | \hat N_f^{\rm \,late} | s \rangle =0$
for sufficiently
large $T_0$.  As before, we interpret (\ref{6.16}) in an
expectation value sense.  However in this case if
the fluctuations of the flux operator are small enough,
then each observation of the
net number of fermions flowing through the detector
corresponding to the smoothed operator (\ref{6.10b})
will yield a number close to $Q$.  In this sense,
fermion number violation is seen with each observation.

\vskip .3in
\section{Concluding Remarks and Open Questions}
\label{sec7}

We have shown that spherical solutions to the classical
equations of motion of $SU(2)$ gauge theory which
propagate freely in the far past have nonzero
topological charge.  Furthermore, the topological
charge of such solutions can be developed as a
power series expansion in $g$.  We presented examples
for which $Q$ has a nonzero contribution
at order $g^5$.
In these examples $Q$ is nonzero for $g$-independent initial conditions.
In particular, this suggests that in a corresponding
quantum process, the number of particles in the initial state
is $g$-independent, and need not be large at small $g$.
We showed
that if we couple
quantized chiral fermions to classical gauge
field solutions with nonzero $Q$,  anomalous fermion number violation
will occur.
In a series of
measurements
using the operators of section VI
with the same background gauge field
and the same quantum mechanical fermion state,
the average number of fermions anomalously produced
is given by the topological
charge $Q$.
For the solutions we have discussed, the topological
charge vanishes as a power of $g$.  This
suggests that anomalous fermion number
violation in high  energy scattering processes
may occur at finite order in $g$, rather than
being exponentially suppressed.

We cannot yet  apply our results directly to scattering
of $W$ bosons at energies much higher than
the sphaleron energy because a  number of important
open questions
remain to be considered.  These questions arise
because we have done a classical calculation
in the spherical ansatz for an $SU(2)$
gauge theory without a Higgs field.  Let us consider each of these
idealizations in turn.

First, in everything we have done, we have treated
the gauge field classically, whereas we are ultimately
interested in quantum scattering processes with a
two-particle initial state.
We need to relate our classical results to
quantum mechanical scattering.  One possible
approach is to
relate small amplitude
classical solutions to quantum mechanical coherent
states with small mean particle number.
In this context, it is interesting to note
that the $g$-expansion we have done is equivalent
to an expansion in the amplitude of the initial
pulse.  Therefore, the topological charge at order
$g^5$ is proportional to the fifth power of the
amplitude of the initial pulse, and is not
exponentially small for small amplitude pulses.
This suggests that there may be scattering processes
with few particles in the initial state in which
$Q$ is not exponentially suppressed.
Another way of addressing the problem might be
to relate
the lowest
order term in a semi-classical expansion
of the $S$-matrix for a quantum scattering process
to the solutions of classical equations of
motion for a specified initial configuration.
The quantum lessons which our
classical results are trying to teach us remain to be learned.

Second, we have worked within the spherical ansatz.
This is a reasonable starting point since
it makes the calculation tractable,
and typically the lowest partial waves dominate
scattering processes.
An important open question here
is that of stability.  We have seen no evidence
that the solutions we discuss are unstable
to perturbations within the spherical ansatz.
However, we have not
checked their stability to non-spherical
perturbations.

Finally, we have
worked in $SU(2)$ gauge theory without a Higgs field.  It remains
to be determined how the inclusion of the Higgs
field affects our results.
It is possible
that
in scattering at very high energies the $W$ mass and sphaleron
barrier are irrelevant and our results
are a good approximation to results in the full theory.
However, including
the Higgs field changes the theory in qualitative
ways.  The theory is no longer scale invariant.
The behavior we describe at early and late times ---
pulses propagating without distortion --- will not
occur once the Higgs field is included.
Thus, while our results are likely applicable
in the full theory in the limit of infinite
energy scattering, it remains to be seen how they can be applied
at high but finite energy.

\acknowledgements

We would like to thank S. Coleman, J. Goldstone, T. Gould,
S. Hsu, K. Johnson,
V. Rubakov, and L. Yaffe for very helpful discussions.

\eject

\vskip .3in
\appendix
\section{Abelian Topological Charge}

Here we show that for a localized solution to Maxwell's equations in empty
space, the topological charge vanishes.
We do not use the spherical ansatz.
It is convenient to work with the
gauge invariant electric and magnetic fields
$\vec{E}(\vec{x},t)$
and
$\vec{B}(\vec{x},t)$
which obey
\begin{mathletters}
\begin{eqnarray}
\vec{\nabla} \cdot \vec{E} = 0
&\hskip.8in&
\vec{\nabla} \times \vec{E} = -\frac{\partial \vec{B}}{\partial t} \\
\vec{\nabla} \cdot \vec{B} = 0
&\hskip.8in&
\vec{\nabla} \times \vec{B} = \frac{\partial \vec{E}}{\partial t}\ .
\end{eqnarray}
\end{mathletters}%
The most general solution can be expressed as
\begin{mathletters}%
\begin{equation}
\vec{E}(\vec{x},t)
= \int d^3 k
\left[
\vec{\epsilon}\, (\vec{k}) \,
\exp{\left(i\omega t - i\vec k \cdot \vec x \right) }
+
\vec{\epsilon}^{~*} (\vec{k}) \, \exp{\left(-i\omega t + i\vec k \cdot \vec x
\right) }
\right]
\end{equation}
and
\begin{equation}
\vec{B}(\vec{x},t)
= \int \frac{d^3 k}{\omega}
\left[
 \vec{k} \times \vec{\epsilon}\, (\vec{k})  \,
\exp{\left(i\omega t - i\vec k \cdot \vec x \right) }
+
 \vec{k} \times \vec{\epsilon}^{~*} (\vec{k})  \,
\exp{\left(-i\omega t + i \vec k \cdot \vec x \right) }
\right] \ ,
\label{A2b}
\end{equation}
\end{mathletters}%
where $\omega=|\vec{k}|$ and
$\vec{\epsilon}\, (\vec{k})$ is an arbitrary function of $\vec{k}$ subject to
$\vec{k} \cdot \vec{\epsilon} \, (\vec{k}) = 0$.

Now the Abelian topological charge can be defined as
\begin{equation}
Q^{\rm Abelian} = \frac{1}{8\pi^2} \int d^4x ~ \vec{E} \cdot \vec{B} \ ,
\label{A3}
\end{equation}
which in terms of $\vec{E}$ and $\vec{B}$ given above is seen to be
\begin{equation}
Q^{\rm Abelian} = -2 \pi^2
\int\limits_0^\infty \omega \, d\omega
\int d\cos\theta \, d\varphi
\delta(2\omega)
\left[
\vec{\epsilon}\, (\vec{k})
\cdot
\left(
\vec{k} \times \vec{\epsilon}\, (-\vec{k}) \right)
+ {\rm c.c.} \right]
{}~.
\end{equation}
The delta function in $\omega$ comes from the $\int dt$ in (\ref{A3}).
Because of this delta function, the integral will vanish unless
$\vec{\epsilon}\, (\vec{k})$ goes like $1/\omega$ as $\omega \to 0$.
If we know $\vec{E}$ and $\vec{B}$ at one time, say $t=0$, we can determine
$\vec{\epsilon}\, (\vec{k})$:
\begin{equation}
\vec{\epsilon}\, (\vec{k}) = \frac{1}{2} \int \frac{d^3 x}{(2\pi)^3}\,
\left[
\vec{E} (\vec{x},0) - \frac{\vec{k}}{\omega} \times \vec{B} (\vec{x},0)
\right] \exp{i \vec{k} \cdot \vec{x}} \ .
\label{A5}
\end{equation}
As $\vec{k}$ goes to zero,
$\vec{\epsilon}\, (\vec{k})$ will not blow up if
\begin{mathletters}
\label{A6}
\begin{equation}
\int d^3x ~ \vec{E} (\vec{x},0) < \infty
\end{equation}
and
\begin{equation}
\int d^3x ~ \vec{B} (\vec{x},0) < \infty\ .
\end{equation}
\end{mathletters}%
For localized pulses these conditions are satisfied and the Abelian
topological charge vanishes.  Note however that finite energy does not ensure
(\ref{A6}) so vanishing Abelian topological charge is not guaranteed under the
most general circumstances.

\section{Evaluating the Fifth Order Topological Charge}

In this appendix we derive equation (\ref{5.9}) for the fifth order topological
charge for initial profiles for which $\gamma_p=0$
and $\alpha_p$ is independent of $g$.
As was explained in the text, this
implies that $\psi_n=0$ for $n$ odd and $\delta_n=0$ for $n$ even.
We first need the third
order contribution to the integrands of equation (\ref{5.4}):
\begin{equation}
  \bigg({ \delta \, \partial_\mu\psi  \over 1 - 2 g \delta}\bigg)^{[3]} =
g^3 \Bigl( \delta_3 \, \partial_\mu \psi_0  +
\delta_1  \partial_\mu\psi_2  + 2 \delta_1^2 \,
\partial_\mu\psi_0 \Bigr) \ .
\label{B1}
\end{equation}
{}From (\ref{5.8}) we see that as $u\to \infty$, $\delta_1 \to -(1/2)
G'(v)^2$ and as $v\to -\infty$, $\delta_1 \to -(1/2)G'(u)^2$.
Therefore, the
last  term in (\ref{B1}) makes
cancelling contributions to the two terms of (\ref{5.4}).
Furthermore, there is no
contribution from the first two terms of (\ref{B1})
as $v\rightarrow -\infty$ since $\delta_3$
and $\psi_2$ vanish in this limit. Thus,
\begin{equation}
   Q^{[5]} = {g^5 \over \pi}\int_ {-\infty}^\infty dv ~ (-\psi_0  \partial_v
\delta_3+
  \delta_1^{{\rm hom}}\partial_v\psi_2  )\bigg\vert_{u=\infty},
\label{B2}
\end{equation}
where the first term has been integrated by parts, and in the second term
we have  used the fact that at large $u$
we can replace $\delta_1$ by the homogeneous
solution $\delta_1^{{\rm hom}}$ given by (\ref{5.7}).

These  single integrals may be turned into double integrals using
\begin{equation}
  \int_{-\infty}^\infty dv \, \eta \, \partial_v \Phi \bigg\vert_{u=\infty} =
 \int_{-\infty}^\infty du \int_{-\infty}^u dv \, \eta \, I \ ,
\label{B3}
\end{equation}
where $\Phi$, given by (\ref{4.8}), is the solution to
equation (\ref{4.7}) with source $I$, and
$\eta$ is
any homogeneous solution of (\ref{4.7}). This relation can be established
by substituting (\ref{4.8}) into the left hand side of (\ref{B3}) and writing
the homogeneous
solution $\eta$ as in equation (\ref{4.3}).

Using (\ref{B3}) we write (B2) as
\begin{equation}
Q^{[5]} = {g^5 \over \pi} \int_{-\infty}^\infty du \int_{-\infty}^u dv ~
\big(-\psi_0 I^\delta_3 + \delta_1^{{\rm hom}}I^\psi_2 \big) \ .
\label{B4}
\end{equation}
Using the equations of motion, (\ref{4.6c})
can be written
\begin{equation}
  I^\psi_2 =  {1\over 2}
\partial_\mu\psi_0\, \partial^\mu\delta_1
+ {4\psi_0\delta_1 \over (u-v)^2} \ .
\label{B5}
\end{equation}
Similarly,
\begin{equation}
 I^\delta_3 =
{1\over 4}
\partial_\mu\delta_1\, \partial^\mu\delta_1
-{1\over 2} \delta_1\,
\partial_\mu\psi_0\, \partial^\mu\psi_0
-{1\over 2} \partial_\mu\psi_0\, \partial^\mu\psi_2
+ {4\delta_1^2 \over (u-v)^2}\ .
\label{B6}
\end{equation}
Using equations (\ref{4.3a}), (\ref{5.7}), (\ref{5.8}), (\ref{B5}),
and (\ref{B6}) we can evaluate every
piece of (B4) except the part which has the term from (\ref{B6})
involving $\psi_2$.
Integrating this term by parts yields
\begin{equation}
  {1\over 2}\int_{-\infty}^\infty du \int_{-\infty}^u dv~\psi_0\,
  \partial_\mu\psi_0\, \partial^\mu\psi_2 \, =\, \int_{-\infty}^\infty du
\int_{-\infty}^u dv~
  \psi_2 \, \partial_u\partial_v \psi_0^2 -
  {1\over 2}\int_{-\infty}^\infty dv~ \psi_2 \, \partial_v \psi_0^2
\bigg\vert_{u=\infty}\ .
\label{B7}
\end{equation}
Using (\ref{5.7}) and (\ref{B3}), the surface term can be written
\begin{equation}
  - {1\over 2}\int_{-\infty}^\infty dv~ \psi_2 \, \partial_v \psi_0^2
\bigg\vert_{u=\infty} ~=~
 \int_{-\infty}^\infty dv~ \psi_2 \, \partial_v \delta_1^{{\rm hom}}
\bigg\vert_{u=\infty}  ~=~
 -\int_{-\infty}^\infty du \int_{-\infty}^u dv~ \delta_1^{{\rm hom}}
 I^\psi_2 \ .
 \label{B8}
\end{equation}
Note that this term  cancels the second term of (\ref{B4}). We now examine the
first
term on the right hand side
of (\ref{B7}). Using (\ref{4.8}) we find
\begin{eqnarray}
   \int_{-\infty}^\infty du \int_{-\infty}^u dv~ \psi_2\, \partial_u\partial_v
\psi_0^2 &=&
  \int_{-\infty}^\infty du \int_{-\infty}^u dv \int_v^u du' \int_{-\infty}^v
dv'~
  F(u,v;u',v') I^\psi_2(u',v') \nonumber\\
& & \qquad\qquad\qquad\qquad\qquad\qquad\qquad
 \partial_u\partial_v \psi_0^2(u,v) \nonumber\\
&=&
  \int_{-\infty}^\infty du' \int_{-\infty}^{u'} dv' \int_{u'}^\infty
du\int_{v'}^{u'} dv~
  F(u,v;u',v')\partial_u\partial_v \psi_0^2(u,v)\nonumber\\
& & \qquad\qquad\qquad\qquad\qquad\qquad\qquad I^\psi_2(u',v')\ .
\label{B9}
\end{eqnarray}
We perform the inner $u$-$v$ integrals as follows. Define the function
\begin{equation}
 \Theta(u',v') = \int_{u'}^\infty du\int_{v'}^{u'} dv~
 F(u,v;u',v')\partial_u\partial_v \psi_0^2(u,v)\ .
\label{B10}
\end{equation}
One can show that
\begin{equation}
  \bigg[\partial_u\partial_v + {2\over (u-v)^2 }\bigg]\Theta(u,v)=
\partial_u\partial_v \psi_0^2=
-{1\over 2}\partial_\mu(\psi_0\, \partial^\mu\psi_0)\ .
\label{B11}
\end{equation}
Hence $\Theta(u,v)$ is a solution to the inhomogeneous equation (\ref{B11})
with
the boundary condition
\begin{equation}
  \lim_{u \rightarrow \infty} \Theta(u,v) = 0 \ .
\label{B12}
\end{equation}
Note that the right hand side of (\ref{B11}) is the same as the right hand
side of
(\ref{4.6a}) with $\delta_0=-\psi_0$.
Also, the solution $\psi_1$ given by (\ref{4.10a})
satisfies the condition (\ref{B12}).
Therefore, $\Theta(u,v)$ is given by (\ref{4.10a}) with $H=-G$:
\begin{eqnarray}
  \Theta(u,v) &=& 2 G'(u)G'(v) + {6\over (u-v)^2}[G(u)-G(v)]^2 \nonumber\\
& & - {4 \over (u-v)^2}[G'(u)+G'(v)][G(u)-G(v)]\ .
\label{B13}
\end{eqnarray}
Thus,
\begin{equation}
 \int_{-\infty}^\infty du \int_{-\infty}^u dv~ \psi_2\,
\partial_u\partial_v \psi_0^2 =
  \int_{-\infty}^\infty du' \int_{-\infty}^{u'} dv'~ \Theta(u',v')
I^\psi_2(u',v') \ .
\label{B14}
\end{equation}
Using (\ref{B6}), (\ref{B7}), (\ref{B8}), (\ref{B13}), and (\ref{B14})
in (\ref{B4}) and integrating by parts several
times we find
\begin{equation}
  Q^{[5]} = {g^5 \over \pi} \int_{-\infty}^\infty du dv~ \bigg[ {1\over
2}\Theta I^\psi_2
 -{3\over 2} {\psi_0 \delta_1 \over (u-v)^2} - {\psi_0^3 \delta_1 \over
(u-v)^2}
 - {1\over 8}{\psi_0^5 \over (u-v)^2} \bigg] \ ,
\label{B15}
\end{equation}
which is equation (\ref{5.9}).
In (\ref{B15}) we found it convenient to use the fact that the integrand
is symmetric
under the interchange of $u$ and $v$ to extend the integration range over the
entire $u$-$v$ plane.
With equation (\ref{B15}) in hand, we can choose $G(x)$, obtain
$\psi_0$ from (\ref{4.3a}), $\delta_1$ from (\ref{5.8}),
$I^\psi_2$ from (\ref{B5}), and $\Theta$ from (\ref{B13}),
and then evaluate the fifth order topological charge.

\end{document}